\global\def\draftcontrol{0}

   \def\versionno{ bulk gubser et.al }

\catcode`\@=11

\expandafter\ifx\csname draftcontrol\endcsname\relax\global\def\draftcontrol{0}
\fi

{\count255=\time\divide\count255 by 60
\xdef\hourmin{\number\count255}
\multiply\count255 by-60\advance\count255 by\time
\xdef\hourmin{\hourmin:\ifnum\count255<10 0\fi\the\count255}}
\def\draftdate{\number\month/\number\day/\number\year\ \ \ \hourmin }

\newcommand\makepapertitle{\par
  \begingroup
    \renewcommand\thefootnote{\@fnsymbol\c@footnote}%
    \def\@makefnmark{\rlap{\@textsuperscript{\normalfont\@thefnmark}}}%
    \long\def\@makefntext##1{\parindent 1em\noindent
            \hb@xt@1.8em{%
                \hss\@textsuperscript{\normalfont\@thefnmark}}##1}%
     \newpage
     \global\@topnum\z@   
     \@makepapertitle
     \thispagestyle{empty}\@thanks
  \endgroup
  \setcounter{footnote}{0}%
  \global\let\thanks\relax
  \global\let\makepapertitle\relax
  \global\let\@makepapertitle\relax
  \global\let\@thanks\@empty
  \global\let\@author\@empty
  \global\let\@date\@empty
  \global\let\@title\@empty
  \global\let\title\relax
  \global\let\author\relax
  \global\let\date\relax
  \global\let\and\relax
  \def\version{\let\version\@version\@gobble}
}
\def\@makepapertitle{%
  \newpage
   \ifnum\draftcontrol=1 {}
   \version\versionno
   \vskip 3em%
   \else
   \hfill\hbox to 3cm {\parbox{4cm}{\@pubnum}\hss}%
   \vskip 3em%
   \fi
   \begin{center}%
   \let \footnote \thanks
     {\LARGE {\@title}}%
     \vskip 1.5em%
     {\normalsize
       \lineskip .5em%
       \begin{tabular}[t]{c}%
         \@author
       \end{tabular}\par}%
     \vskip 1.5em%
     {\@bstract}%
     \end{center}%
     \vskip 1.5em
     \@date%
   \par
}

\gdef\@pubnum{}
\def\pubnum#1{%
  \gdef\@pubnum{#1}}

\gdef\@bstract{}
\def\Abstract#1{%
  \gdef\@bstract{%
   \parbox{\textwidth-0pc}{%
   \centerline{\bf Abstract}\penalty1000%
\kern.2cm%
\noindent
\renewcommand\baselinestretch{1.0}%
{#1}}}
}

\def\ps@paper{\let\@mkboth\@gobbletwo%
     \ifnum\draftcontrol=1
    \def\@oddfoot{\hbox to \textwidth{\tiny \versionno \hfil\tiny\draftdate}%
    \hskip -\textwidth \hbox to \textwidth{\hfil\rm\thepage\hfil}}%
     \else\def\@oddfoot{\hbox to \textwidth{\hfil\rm\thepage\hfil}}
     \fi
     \let\@evenfoot\@oddfoot
}

\def\body{\clearpage
          \pagestyle{paper}
    }

\def\@version#1{\ifnum\draftcontrol=1
\typeout{}\typeout{#1}\typeout{}
\vskip3mm\centerline{\hbox{\fbox{\normalsize{\tt DRAFT -- #1 -- }
                   {\draftdate}}}}\vskip3mm
\fi}
\let\version\@version
\long\def\eqlabel#1{\ifnum\draftcontrol=1
                    \tag@false  
                    \tag*{(\theequation) \hbox to -0.2cm{\hspace{0cm}\small{#1}\hss}}
                    \refstepcounter{equation}
                    \edef\@currentlabel{\theequation}
                    \ltx@label{#1}          
                    \else
                    \label{#1}
                    \fi
                    }
\let\st@bibitem\@bibitem
\let\st@lbibitem\@lbibitem
\ifnum\draftcontrol=1
  \def\@bibitem#1{%
    \st@bibitem{#1}\a@@label{#1}\ignorespaces}
  \def\@lbibitem[#1]#2{%
    \st@lbibitem[#1]{#2}\a@@label{#2}\ignorespaces}
  \def\a@@label#1{%
    \gdef\a@lab{\smash{\normalfont\small#1}}
    \ifvmode
      \if@inlabel
        \global\setbox\@labels\hbox{%
          \llap{\a@lab\let\a@lab\relax
                \kern\@totalleftmargin\kern\marginparsep}%
          \box\@labels}%
      \fi
    \fi}
\fi

\documentclass[12pt,letterpaper]{article}

\usepackage{amsmath,amssymb,array,calc,epsfig,rotating,bm}
\usepackage[sort]{cite}
\usepackage{graphicx}
\usepackage{psfrag,verbatim}
\usepackage[hypertex]{hyperref}

\def\hri#1#2{\href{http://arxiv.org/abs/#1}{[#1]#2}}
\def\hre#1#2{\href{http://arxiv.org/abs/#1/#2}{[ArXiv:#1/#2]}}

\ifnum\draftcontrol=1
\fi

\renewcommand\baselinestretch{1.25}
\renewcommand\section{\@startsection {section}{1}{\z@}%
                                   {-3.5ex \@plus -1ex \@minus -.2ex}%
                                   {2.3ex \@plus.2ex}%
                                   {\normalfont\large\bfseries}}
\renewcommand\subsection{\@startsection{subsection}{2}{\z@}%
                                   {-3.25ex\@plus -1ex \@minus -.2ex}%
                                   {1.5ex \@plus .2ex}%
                                   {\normalfont\normalsize\bfseries}}
\renewcommand\subsubsection{\@startsection{subsubsection}{3}{\z@}%
                                   {-3.25ex\@plus -1ex \@minus -.2ex}%
                                   {1.5ex \@plus .2ex}%
                                   {\normalfont\normalsize\it}}
\renewcommand\paragraph{\@startsection{paragraph}{4}{\z@}%
                                   {-3.25ex\@plus -1ex \@minus -.2ex}%
                                   {1.5ex \@plus .2ex}%
                                   {\normalfont\normalsize\bf}}

\numberwithin{equation}{section}

\def\revise#1       {\raisebox{-0em}{\rule{3pt}{1em}}%
                     \marginpar{\raisebox{.5em}{\vrule width3pt\
                     \vrule width0pt height 0pt depth0.5em
                     \hbox to 0cm{\hspace{0cm}{%
                     \parbox[t]{4em}{\raggedright\footnotesize{#1}}}\hss}}}}

\def\cala         {{\cal A}}

\def\calh         {{\cal H}}

\def\calm         {{\cal M}}
\def\caln         {{\cal N}}
\def\calo         {{\cal O}}
\def\calp         {{\cal P}}

\def\calv         {{\cal V}}

\def\del          {\partial}

\def\ee           {{\rm e}}

\def\tr           {\mathop{\rm Tr}}

\def\Im           {{\rm Im\hskip0.1em}}

\def\sqr#1#2{{\vcenter{\vbox{\hrule height.#2pt
 \hbox{\vrule width.#2pt height#1pt \kern#1pt
 \vrule width.#2pt}\hrule height.#2pt}}}}

\def\labb{\eqlabel}
\def\6{\partial}
\def\f{\phi}
\def\a{\alpha}

\def\le{\left}
\def\ri{\right}

\def\C0{{\bf C_0}}
\def\Y0{{\bf Y_0}}
\def\G0{{\bf G_0}}

\def\sq
\def\a{\alpha}

\def\l{\lambda}

\def\tr{{\rm Tr}}

\def\o{\omega}

\def\bz{\begin{itemize}}
\def\ez{\end{itemize}}
\def\bn{\begin{enumerate}}
\def\en{\end{enumerate}}

\def\be{\begin{equation}}
\def\ee{\end{equation}}
\def\bea{\begin{eqnarray}}
\def\eea{\end{eqnarray}}

\newcommand{\ft}[2]{{\textstyle{\frac{#1}{#2}}}}

\def\a{\alpha}

\newcommand{\beq}{\begin{equation}}
\newcommand{\eeq}{\end{equation}}
\newcommand{\beqa}{\begin{eqnarray}}
\newcommand{\eeqa}{\end{eqnarray}}
\newcommand{\beqar}{\begin{eqnarray*}}
\newcommand{\eeqar}{\end{eqnarray*}}

\renewcommand{\eqref}[1]{(\ref{#1})}

\newcommand{\ie}{{\it i.e.,}\ }

\def\a{\alpha}
\def\w{\omega}

\def\dd{{\delta}}
\def\l{\lambda}

\catcode`\@=12

\begin{document}


\title{\bf Holographic bulk viscosity: GPR vs EO}
\pubnum
{UWO-TH-11/5\\ CCTP-2011-12\\CERN-PH-TH-2011-074}

\date{April 2011}

\author{\\[0.5cm]\bf\large
Alex Buchel$^{1,2}$, Umut G\"ursoy$^3$, Elias Kiritsis$^{4,5}$\\[.4cm]
$^1$ \it Department of Applied Mathematics, University of Western Ontario,\\ \it London, Ontario N6A 5B7, Canada\\~\\
$^2$\it Perimeter Institute for Theoretical Physics,\it  Waterloo, Ontario N2J 2W9, Canada\\~\\
$^3$ \it Theory Group, Physics Department, CERN, \it CH-1211 Geneva 23, Switzerland\\~\\
$^4$ \it \href{http://hep.physics.uoc.gr/}{Crete Center for Theoretical Physics}, Department of Physics, University of Crete\\
\it 71003 Heraklion, Greece\\~\\
$^5$ \it \href{http://www.apc.univ-paris7.fr/APC_CS/}{Laboratoire APC,
Universit\'e Paris-Diderot Paris 7, CNRS UMR 7164}, \\
\it 10 rue Alice Domon et L\'eonie Duquet, 75205 Paris Cedex 13, France.\\[.2
cm]}

\Abstract{Recently Eling and Oz (EO) proposed a formula for the holographic bulk
viscosity, in \href{http://arxiv.org/abs/1103.1657}{arXiv:1103.1657},
derived from the null horizon focusing equation. This formula seems
different from that obtained earlier by Gubser, Pufu and Rocha (GPR)
in \href{http://arxiv.org/abs/0806.0407}{arXiv:0806.0407} calculated
from the IR limit of the two-point function of the trace of the stress
tensor. The two were shown to agree only for some simple scaling
cases.  We point out that  the two formulae agree in two
non-trivial holographic theories describing RG flows.  The first is
the strongly coupled $\caln=2^*$ gauge theory plasma. The second is
the semi-phenomenological model of Improved Holographic QCD.}

\makepapertitle

\body

\version\versionno
\tableofcontents

\section{Introduction and Summary}

The bulk viscosity  of strongly coupled thermal systems is a quantity of phenomenological importance. On the other hand it is quite difficult to compute.
The main difficulty arises from the fact that the bulk viscosity, as one of the plasma deep-infrared transport
coefficients, is sensitive to the microscopic (ultraviolet) parameters
of the theory\footnote{The only exception is a conformal theory, where the scale invariance imposed the bulk viscosity to vanish.}.
 Necessarily, a computation of the bulk viscosity in
a given system requires the understanding of its physics over a wide range of scales.
It is perhaps not surprising that the first computation of the bulk viscosity in gauge theory plasmas \cite{v3} was
performed in the framework of gauge  theory/string theory correspondence \cite{juan,adscft}.

In \cite{n21} it was observed that for a large class of holographic models,
the bulk viscosity of the strongly coupled plasma satisfied the following bound
\begin{equation}
\frac \zeta\eta\ge 2 \left(\frac 13 -c_s^2\right)\,,
\eqlabel{bound}
\end{equation}
where $\eta$ is the universal shear viscosity of strongly coupled holographic plasma \cite{u1,u2,u3,u4},
 and $c_s$ is the speed of sound waves in plasma.
The computation of the bulk viscosity which led to \eqref{bound} was based on analyzing the dispersion
relation of the sound waves in plasma. Alternatively, the bulk viscosity can be computed using the
Kubo formula
\begin{equation}
\zeta=-\frac 49 \lim_{\w\to 0}\ \frac 1\w\ \Im G_R(\w)\,,
\eqlabel{kubo}
\end{equation}
where $G_R$ is the retarded correlation function of the stress-energy tensor
\begin{equation}
G_R(\w)=-i \int dt d^3 x e^{i\w t }\Theta(t)\langle [\frac 12 T_i^{\ i}(t,\vec{x}),\ \frac 12 T_k^{\ k}(0,0)]\rangle\,.
\eqlabel{gr}
\end{equation}
The holographic computations of the correlator \eqref{gr} for a certain class of dual gravitational models by Gubser, Pufu and Rocha (GPR)
was reported in \cite{f1}. It was claimed that some of the Einstein-scalar models considered  led to a violation of the bound
\eqref{bound}. On the other hand, in the Improved Holographic QCD model, \cite{GK}, the bound (\ref{bound}) is comfortably obeyed, \cite{transport}.

Recently an alternative expression for the bulk viscosity in strongly coupled plasmas with a holographic dual
was obtained by Eling and Oz (EO) in  \cite{eo}. They have analyzed directly the hydrodynamic limit of the equations of motion of a generic Einstein-scalar theory
and derived a formula for the bulk viscosity that is apparently different from the GPR formula.
The EO formula is very general and reads
\be
{\zeta\over \eta}=\sum_i\left[s{\partial \phi^i_{h}\over \partial s}+\sum_a\rho_a{\partial \phi^i_h\over \partial \rho^a}\right]
\ee
where $i$ labels different bulk scalars, $\phi^i_h$ is the value of the i-th scalar at the horizon, and $\rho^a$ are different conserved charged densities.
The case comparable with GPR, involves a single scalar field and no charge density.
In \cite{eo} the two formulae were shown to give the same result in cases where the adiabatic approximation to the equations is valid, but their equivalence in more general cases
was put in doubt.

In this paper we re-analyze the bulk viscosity in two non-trivial holographic theories, the  bosonic  $\caln=2^*$ theory \cite{ns1} as well as Improved Holographic QCD, \cite{GK,GKMN1}.
The bosonic $\caln=2^*$ theory is $N=4$ superYM, with a non-trivial (and equal) mass for 4 of the 6 scalars.
Improved holographic QCD on the other hand is a semi-phenomenological Einstein-scalar theory tuned to match non-supersymmetric Yang Mills theory in the large $N_c$ limit\footnote{Similar
Einstein-scalar theories were also proposed to describe the crossover behavior of QCD with light quarks in \cite{gu}.} .

We point out that the analysis done in \cite{f1},
when applied to $\caln=2^*$ gauge theory plasma
\cite{ns1,ns2,ns3,ns4,ns5,ns6}
at high temperatures, agrees with earlier computations reported in   \cite{v3,ya,eo}. The agreement is also checked numerically at all temperatures.

Both in the $\caln=2^*$ theory and Improved Holographic QCD we confirm the agreement between GPR and EO formulae
 for the
holographic bulk viscosity.

\section{The GPR formula for the holographic bulk viscosity}
Here we mostly follow \cite{f1}. For details we refer to the original work.

Consider a gravitational system, dual to some strongly
coupled gauge theory plasma, described by an Einstein-dilaton system of the form
\begin{equation}
S=\frac{1}{16\pi G_5} \int d^5x\sqrt{-g}\left[R-\frac 12 \left(\del\phi\right)^2-V(\phi)\right]\,.
\eqlabel{5dact}
\end{equation}
The black brane background geometry dual to a thermal state of the plasma takes the form
\begin{equation}
ds^2\equiv g_{\mu\nu}^{(0)}dx^\mu dx^\nu=e^{2A(r)}\left[-h(r) dt^2 +d\vec{x}^2\right]+e^{2B(r)}\frac{dr^2}{h(r)}\,,\qquad \phi=r\,.
\eqlabel{bac}
\end{equation}
Notice that the field $\phi$ was chosen as a radial coordinate\footnote{One might worry
whether $\phi$ is monotonic from the boundary to the black brane horizon. In the Einstein-dilaton theory, there are solutions where $\phi'$ vanishes along the flow.
These where analyzed in \cite{GKMN2} and shown to be unphysical, violating the Gubser bound \cite{gb}. Therefore, this is not expected to happen in the middle of an RG flow.
At theories with an extra gauge field and at finite density however, it is possible that $\phi'=0$ in a physical solution. A class of such examples were studied recently in \cite{Char}.}.

The background equations of motion take a simple form
\begin{equation}
\begin{split}
0=&A'' -A'B' +\frac  16\,,\\
0=&h''+(4 A'-B') h'\,,\\
0=& 6 A' h' +h (24 A'^2-1)+2e^{2B}V\,,\\
0=&4 A'-B'+\frac {h'}{h}-\frac{e^{2B}}{h}V'\,.
\end{split}
\eqlabel{beom}
\end{equation}

To compute the correlation function \eqref{gr}, the authors of \cite{f1} considered an $SO(3)$-invariant fluctuation of the
metric $\dd g_{\mu\nu}(t,\phi)\to e^{-i \w t} g^{(0)}_{\mu\nu}(\phi) H_{\mu\nu}(\phi)$
in the gauge $\dd \phi=0$. It was shown that the equation for $H_{11}$ decouples from the rest of the fluctuation equations and is
\footnote{We independently reproduced this equation. We also verified the consistency of the
gauge choice $\dd\phi=0$.}
\begin{equation}
H_{11}''=\left(-\frac{1}{3A'}-4 A'+3 B'-\frac{h'}{h}\right)H'_{11}+\left(-\frac{e^{-2 A+2 B}}{h^2}\w^2+
\frac{h'}{6h A'}-\frac{h'B'}{h}\right)H_{11}\,.
\eqlabel{h11eq}
\end{equation}
One  further has to solve \eqref{h11eq} with the following UV ($r\to 0$) and IR ($r\to \phi_h$)\footnote{Note that in the gauge we are working the position of the black hole horizon $r_h$
is identified with the value of the scalar at the horizon $\phi_h$.}
boundary conditions:
\begin{equation}
{\rm UV}:\qquad \lim_{r\to 0} H_{11}=1\,,
\eqlabel{uv}
\end{equation}
\begin{equation}
{\rm IR}:\qquad  H_{11}\to c_{11}^- (\phi_h-r)^{-i \w /4\pi T}+ 0\times  (\phi_h-r)^{+i \w /4\pi T}\,,\qquad
{\rm as}\qquad r\to \phi_h \,.
\eqlabel{ir}
\end{equation}
The bulk viscosity, computed from \eqref{kubo}, is given by \cite{f1}
\begin{equation}
\frac{\zeta}{\eta}=\frac {1}{9 A'(\phi_h)^2}\ \lim_{\w\to 0} |c_{11}^-|^2\,,
\eqlabel{bulk}
\end{equation}
where one has to use the universality of the shear viscosity \cite{u1,u2,u3,u4}.
The authors of \cite{f1} used \eqref{beom} to obtain
\begin{equation}
A'(\phi_h)=-\frac{V(\phi_h)}{3 V'(\phi_h)}\,,
\eqlabel{apg}
\end{equation}
and arrived at the final formula for the bulk viscosity ratio
\begin{equation}
\frac{\zeta}{\eta}=\frac {V'(\phi_h)^2}{V(\phi_h)^2}\ \lim_{\w\to 0} |c_{11}^-|^2\,.
\eqlabel{bulkf}
\end{equation}

\subsection{The EO versus GPR formula for the bulk viscosity}

In \cite{eo} Eling and Oz, by analyzing the hydrodynamic limit of the scalar-tensor equations,  produced the following   expression for the holographic bulk viscosity\footnote{
The formula derived in \cite{eo} applies also to systems at finite charge density. Here we restrict our attention to zero charge density systems.
 The EO formula was further tested in\cite{beo}.}
\begin{equation}
\frac{\zeta}{\eta}\bigg|_{EO}=\left(s\ \frac{d\phi_h}{ds}\right)^2=\frac{1}{9A'(\phi_h)^2}\,,
\eqlabel{eobulk}
\end{equation}
Even though (up to a factor of $c_{11}^-$) \eqref{eobulk} and \eqref{bulk} appear to be the same,
they are, in fact, {\it different}: in \eqref{eobulk},
\begin{equation}
A'(\phi_h)\bigg|_{EO}=\frac{d\ \left(\lim_{\phi\to \phi_h} A(\phi)\right)}{d \phi_h}\ne
A'(\phi_h)\bigg|_{GPR}=\lim_{\phi\to \phi_h} \frac{dA(\phi)}{d\phi}=\lim_{\phi\to \phi_h} -\frac{V(\phi)}{3V'(\phi)}\,.
\eqlabel{corder}
\end{equation}

To be specific, in $\caln=2^*$ gauge theory plasma at high temperature (see Appendix \ref{A} for some details)
\begin{equation}
A'(\phi_h)\bigg|_{EO}=\frac{\pi T^2\sqrt{6}}{m_b^2}+\calo\left(\left(\frac{m_b^2}{T^2}\right)^0\right)\,,
\eqlabel{apeo}
\end{equation}
\begin{equation}
A'(\phi_h)\bigg|_{GPR}=\frac{2 \pi T^2\sqrt{6}}{m_b^2}+\calo\left(\left(\frac{m_b^2}{T^2}\right)^0\right)\,.
\eqlabel{apgu}
\end{equation}

From \eqref{apeo} and $\eqref{apgu}$ it is clear that \eqref{bulk} would produce the correct expression for the
$\caln=2^*$ plasma bulk viscosity, provided\footnote{We assume the  $\w\to 0$ limit taken.}
\begin{equation}
|c_{11}^-|\ \bigg|_{\caln=2^*,\ {\rm prediction}}= 2+\calo\left(\frac{m_b^2}{T^2}\right)\,.
\eqlabel{c11pred}
\end{equation}
In the next section we explicitly compute $c_{11}^-$, and find that it agrees with \eqref{c11pred}.

\section{Bulk viscosity calculation in the $\caln=2^*$ plasma}

In this section we will first address the calculation of bulk viscosity in the $\caln=2^*$ theory.

\subsection{The computation of $c_{11}^-$ in $x$-gauge}
We find it convenient to recast the equation for $H_{11}$ in terms of
$x$ coordinate, defined as
\begin{equation}
x\equiv 1- \sqrt{h}\,.
\end{equation}
Notice that $x\to 0_+$ corresponds to the boundary and $x\to 1_-$ to the horizon.

In this gauge the background equations take the form (all the derivatives are with respect to $x$):
\begin{equation}
0=A''-4 (A')^2+\frac{A'}{1-x}+\frac 16 (\phi')^2\,,
\eqlabel{bac1}
\end{equation}
\begin{equation}
0=\phi''-\frac{V_{,\phi}}{2 V}\ (\phi')^2+\frac{\phi'}{x-1}+\frac{6V_{,\phi}\ A'(2 A'(x-1)+1)}{V\ (1-x)}\,,
\eqlabel{bac2}
\end{equation}
where
\begin{equation}
V_{,\phi}\equiv\frac{d V}{d \phi}\,.
\eqlabel{vpdef}
\end{equation}
The equation for $H_{11}$ is somewhat complicated
\begin{equation}
0=H_{11}''+\calh_1\ H_{11}'+\calh_2\ H_{11}\,,
\eqlabel{hx}
\end{equation}
where we collected the coefficients $\calh_i$ in Appendix \ref{B}.
In order to compute the bulk viscosity \eqref{bulk}, we need to solve \eqref{hx}
subject to the following boundary conditions:
\begin{equation}
{\rm UV}:\qquad \lim_{x\to 0_+} H_{11}=1\,,
\eqlabel{uv1}
\end{equation}
\begin{equation}
{\rm IR}:\qquad  H_{11}\to \tilde{c}_{11}^- (1-x)^{-i \w /2\pi T}+ 0\times  (1-x)^{+i \w /2\pi T}\,,\qquad
{\rm as}\qquad x\to 1_- \,.
\eqlabel{ir1}
\end{equation}
For generic $\omega$, $\tilde{c}_{11}^-\ \ne\ {c}_{11}^-$ (see \eqref{ir}), however, in the hydrodynamic limit
\begin{equation}
\lim_{\w\to 0}\tilde{c}_{11}^-=\lim_{\w\to 0} {c}_{11}^-\ \equiv c_{11}^-\,.
\eqlabel{hydro}
\end{equation}

We can test \eqref{bac1}-\eqref{hx} with a simple, exactly solvable  background, like the exponential potential case. This is described in Appendix \ref{cr}.

\subsection{$\caln=2^*$ plasma at high temperatures}

The effective action of the gravitational dual to strongly coupled $\caln=2^*$ plasma
with a bosonic mass deformation is given by \cite{ns6}
\begin{equation}
\begin{split}
S=&\frac{1}{4\pi G_5}\,
\int_{\calm_5} d\xi^5 \sqrt{-g}\left[\ft14 R-3 (\del\a)^2-
V\right]\,,
\end{split}
\eqlabel{action5}
\end{equation}
where the potential is
\footnote{We set the five-dimensional gauged
supergravity coupling to one. This corresponds to setting the
radius $\ell$ of the five-dimensional sphere in the undeformed metric
to $2$.}
\begin{equation}
V=-\frac 14 e^{-4\alpha}-\frac12 e^{2\alpha}\,.
 \eqlabel{pp}
\end{equation}
Notice that  the canonically normalized scalar is $\phi=\sqrt{24} \a$, and therefore
\begin{equation}
V_{,\phi}=\frac{1}{\sqrt{24}}\ V_{,\a}\,.
\eqlabel{vpva}
\end{equation}

We will study the theory \eqref{action5} in the high-temperature regime.
In this case (see Appendix \ref{A})
\begin{equation}
\begin{split}
&e^\a\equiv \rho=1+\dd_1 \a_1+\calo(\dd_1^2)\,,\\
&A=\ln \dd_3-\frac 14 \ln(2x-x^2)+\dd_1^2 A_1+\calo(\dd_1^4)\,.
\end{split}
\eqlabel{hightb}
\end{equation}
In the hydrodynamic limit, \ie $\w\to 0$, and to leading order in $\dd_1$, we find
\begin{equation}
\begin{split}
0+\calo(\dd_1)=&H_{11}''+\frac{(x \a_1' (2-x) (x^2-2 x+4)-2 \a_1 (1-x)}{\a_1' x^2 (1-x) (2-x)^2}\ H_{11}'
\\
&+\frac{2 ((4 x-2 x^2) \a_1'+\a_1 (x-1))}{(2-x)^2 (1-x)^2 \a_1' x^2}\ H_{11}\,.
\end{split}
\eqlabel{h11n2}
\end{equation}
Notice that there is dependence only on $\a_1$, which satisfied the following equation
\begin{equation}
0=\a_1''+\frac{1}{x-1}\ \a_1'
+\frac{1}{x^2(2-x)^2}\ \a_1\,.
\eqlabel{aeq}
\end{equation}

Even though we know an analytic solution for $\a_1$ (see \eqref{orko1}),
we can not solve for $H_{11}$ analytically. We find it convenient
to  use numerical techniques to solve both \eqref{h11n2} and \eqref{aeq}.
Near the boundary we have
\begin{equation}
\a_1=\sqrt{x}\ \left(\sum_{n=0}^\infty \sum_{k=0}^1 a_{n,k}\ x^n\ln^k x\right)\,,
\eqlabel{uvexp}
\end{equation}
with normalization\footnote{The overall normalization
of $\a_1$ is arbitrary, we choose the leading $\ln x$ coefficient to be 1.
}   $a_{0,1}=1$, and
\begin{equation}
a_{n,k}=a_{n,k}\biggl(a_{0,0}\biggr)\,.
\eqlabel{higher}
\end{equation}
For example, for the first few terms we have:
\begin{equation}
a_{1,0}=\frac 12 +\frac 14 a_{0,0}\,,\qquad a_{1,1}=\frac 14\,,\qquad a_{2,0}=\frac {5}{16}+\frac{5}{32}a_{0,0}\,,\qquad a_{2,1}=\frac{5}{32}\,.
\eqlabel{few}
\end{equation}
The asymptotic expansion for $H_{11}$ is a bit unusual because the perturbing operator has scaling dimension 2:
\begin{equation}
H_{11}=\sum_{n=0}^\infty \sum_{k=0}^{n+1} h_{n,k}\ x^n\ \frac{1}{(a_{0,0}+2+\ln x)^k}\,,
\eqlabel{uvexh}
\end{equation}
with normalization $h_{0,0}=1$, see \eqref{uv1}. Here,
\begin{equation}
h_{n,k}=h_{n,k}\biggl(h_{0,1}\biggr)\,.
\eqlabel{hpar}
\end{equation}
For the first few terms we have:
\begin{equation}
\begin{split}
&h_{1,0}=-1\,,\qquad h_{1,1}=-h_{0,1}\,,\qquad h_{1,2}=-\frac 12 h_{0,1}\,,\qquad\\
&h_{2,0}=-\frac 14\,,\qquad h_{2,1}=\frac 14(1-h_{0,1})\,,\qquad h_{2,2}=\frac{9}{16}h_{0,1}\,,\qquad h_{2,3}=\frac 14 h_{0,1}\,.
\end{split}
\eqlabel{hpara}
\end{equation}

Near the horizon, $y\equiv 1-x$, we obtain
\begin{equation}
\begin{split}
&\a_1=a^h \sum_{n=0}^\infty a_n^h y^{2n}=a^{h}\left(1-\frac 14 y^2-\frac {7}{64}y^4-\frac{17}{256}y^6+\cdots\right)\,,\\
&H_{11}=h^h\sum_{n=0}^\infty h_n^h y^{2n}=h^{h}\left(1-\frac 18 y^2-\frac {3}{64}y^4-\frac{27}{1024}y^6+\cdots\right)\,.
\end{split}
\eqlabel{ah}
\end{equation}

Altogether we have four integration constants:
\[
\{\ a_{0,0}\,,\ h_{0,1}\,,\ a^h\,,\ h^h\ \}\,,
\]
precisely what is needed to solve uniquely the system of two second order ODEs: \eqref{h11n2} and \eqref{aeq}.
Using numerical techniques developed in \cite{abk} we find
\begin{equation}
\begin{split}
&a_{0,0}=-2.079441(5)\,,\qquad  a^h=-2.221441(5)\,,\\
&h_{0,1}=-2.000000(0)\,,\qquad h^h=2.000000(0)\,.
\end{split}
\eqlabel{results}
\end{equation}
Of course, $\{a_{0,0}, a^h\}$ are known analytically from \eqref{orko1},
\[
\{a_{0,0}\,,\ a^h\}=\{-\ln 8\,, -\frac{\pi}{\sqrt{2}}\ \}\,,
\]
and are in excellent agreement with \eqref{results}.

From \eqref{results},
\begin{equation}
c_{11}^-\bigg|_{\caln=2*}=h^h\ = 2\,,
\eqlabel{cn2}
\end{equation}
to a very good accuracy,
confirming the agreement of bulk viscosity for the
high-temperature $\caln=2^*$ plasma from \eqref{bulk} with earlier computations
\cite{v3,ya,eo}.

\begin{figure}[t]
\begin{center}
\psfrag{ratio}{{$\left(\frac{{\zeta}/{\eta}|_{GPR}}{{\zeta}/{\eta}}-1\right)$}}
\psfrag{mb2}{{$\frac{m_b^2}{T^2}$}}
\includegraphics[width=4in]{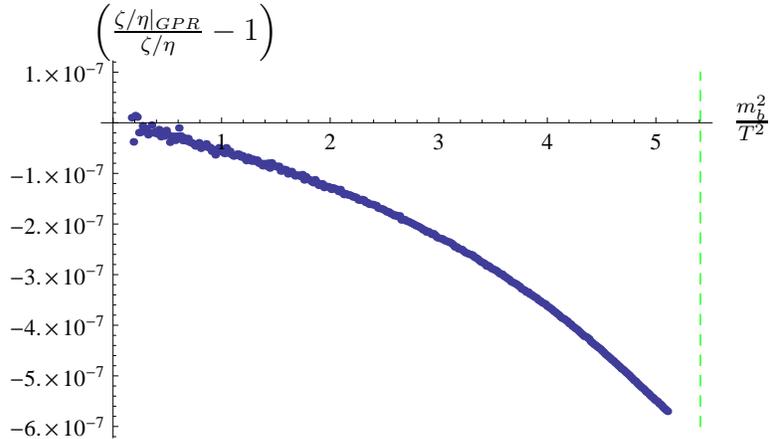}
\end{center}
  \caption{(Color online)
Comparison of the GPR prediction for  $\caln=2^*$ plasma bulk viscosity
with the explicit
computations from the quasinormal modes \cite{n21}.
The dashed vertical green line represents the critical point of the
theory $\frac{m_b^2}{T^2}=\dd_c=5.4098(6)$ associated with the
second-order phase transition \cite{cr2,cr3}.
 } \label{figure1}
\end{figure}

\subsection{$\caln=2^*$ plasma at generic temperatures for
$\frac{m_b^2}{T^2}>0$}

It is straightforward to extend the analysis of the previous section to
generic temperatures $\caln=2^*$ gauge theory plasma for physical mass
deformations, \ie $\frac{m_b^2}{T^2}>0$. The background geometry was studied
in \cite{ns6}, and the bulk viscosity (from the sound waves dispersion
relation) was computed in \cite{n21}. The results of the analysis are
reported in Figure \ref{figure1}. We further verified that the
GPR formula \eqref{bulkf} for the bulk viscosity, when
applied to $\caln=2^*$ plasma, agrees with the
bulk viscosity of the theory at criticality \cite{n21} computed from the sound
waves dispersion relation to $\approx 5\times 10^{-7}$.

\section{Bulk viscosity calculation in Improved Holographic QCD}

In this section we perform an independent calculation of the coefficient $c_{11}^-$ by the methods developed
in \cite{GKMN2, transport}. As described in section 7 of \cite{GKMN2}, one can work out the thermodynamics of gravity-scalar system
entirely by solving a system of coupled first order equations for the so-called {\it phase variables} introduced below.
\subsection{Computation of $c_{11}^-$ using phase variables\label{xxx}}

Starting with the action  (\ref{5dact}) we look for a black-hole solution of the form,
\begin{equation}\label{BH}
  ds^2 = e^{2A(r)}\le( h^{-1}(r)dr^2 + dx_{d-1}^2 + dt^2 h(r) \ri)\,, \qquad \f=
  \f(r)\,.
\end{equation}
We are interested in solutions that are asymptotically AdS. In the dual field theory
this corresponds to the presence of conformal invariance in the UV that is broken either explicitly by a mass deformation as in the ${\cal N}=2^*$ theory
 or by a marginal deformation as in the phenomenological models of \cite{GK}.

In the gauge $\delta\f=0$ one can equivalently use $\f$ as the radial variable. Defining the following phase variables \cite{GK}\cite{GKMN2},
 \begin{equation}\label{XY}
  X(\f)\equiv \frac14 \sqrt{\frac23}\frac{\f'}{A'}\,, \qquad  Y(\f)\equiv
  \frac{1}{4}\frac{h'}{h A'}\,,
\end{equation}
the Einstein's equations can be reduced to
\bea\label{Xeq}
\frac{dX}{d\f} &=& -\sqrt{\frac23}~(1-X^2+Y)\le(1+\sqrt{\frac38}\frac{1}{X}\frac{d\log V}{d\f}\ri)\,,\\
\frac{dY}{d\f} &=& -\sqrt{\frac23}~(1-X^2+Y)\frac{Y}{X }\,. \label{Yeq} \eea
This coupled first order system is sufficient to determine all of the
thermodynamic properties (and dissipation) of the gravitational
theory \cite{GKMN2}. Once a solution to (\ref{Xeq},\ref{Yeq}) is constructed, the metric
functions can be determined as,
\bea
A(\f) &=& A_0 + \frac{1}{4}\sqrt{\frac23}\int_{\f_0}^{\f} \frac{1}{X}d\tilde{\f}\,,\label{Aeq}\\
h(\f) &=&  \exp\le(\sqrt{\frac23} \int_{\f_0}^{\f} \frac{Y}{X}d\tilde{\f}\ri)\,.\label{feq}
\eea
Here $\f_0$ corresponds to the UV value at the boundary corresponding to the UV AdS minimum of the potential.
$A_0$ is an integration constant that essentially determines the energy scale of the breaking of conformal symmetry.

The thermodynamics of the black-hole can directly be determined
as follows. The free energy is given by
\be F(\f_h) = \frac{1}{4G_5} \int^{\infty}_{\f_h}
d\tilde{\f}_h~e^{3A(\f_h)}~\frac{dT}{d\tilde{\f}_h}\,.\label{Feq1}
\ee
These backgrounds satisfy the 1st law of thermodynamics $S =
-dF/dT$. Equation  (\ref{Feq1}) directly follows from integrating this
equation, where
\begin{equation}\label{ent31}
S = \frac{1}{4\pi G_5}e^{3A(\f_h)}
\end{equation}
is used. In the integration in (\ref{Feq1}) one should make sure that the UV asymptotics is kept fixed as
$\f_h$ is varied. This is explained in the case of marginal deformations in section 7 of \cite{GKMN2}.

The temperature as a function of $\f_h$ is obtained from
\be\labb{Teq1} T(\f_h) =
\frac{\ell}{12\pi}~e^{A(\f_h)}~V(\f_h)~e^{\sqrt{\frac23}
\int_{\f_0}^{\f_h} X(\f)~d\f}\,. \ee
Once we solve (\ref{Xeq}) and (\ref{Yeq}) above, we can calculate the free energy as a function of $A_0$ and $T$ by employing the
formulae above.

\subsection{The fluctuation equation}
\label{XYcal}

The fluctuation equation (\ref{h11eq}) in terms of the phase variables read,

\be\label{hwXY}
H_{11}'' = c(\f) H_{11}'+ d(\f) H_{11}\,,\ee
where
\bea\label{cf}
c(\f) &=& \frac{1-X^2+Y}{X}\left(\frac{4}{\sqrt{6}}+\frac{3}{2X}\frac{V'}{V}\right)\,, \\
d(\f) &=& -\frac{2Y}{3X^2}(1-X^2+Y)(1+\frac{\sqrt{3}}{\sqrt{8}X}\frac{V'}{V}) -
\left(\frac23 \frac{\o Y}{4\pi T
X}\right)^2e^{-\sqrt{\frac32}\int^{\f_h}_{\f}\frac{1}{X}}\label{df}\,. \eea

In passing, we note that changing the variable back to the original radial coordinate in (\ref{BH}) produces a rather simple equation
 \cite{G}:
\be
\ddot{H}_{11} +  \dot{H}_{11}\le(3\dot{A} + \frac{\dot{h}}{h} +{\bf
2\frac{\dot{X}}{X}} \ri) + \dot{H}_{11}\le(\frac{\o_h^2}{h^2}-{\bf
\frac{\dot{h}}{h}\frac{\dot{X}}{X}}\ri)=0\,,\labb{hb}\ee
where we emphasized
the new terms in the bulk fluctuation eq. that arise from mixing
of the rotationally invariant graviton excitations and the
dilaton. The {\em normalized} frequency is defined by $\o_h=\o
r_h$.
This equation compares with the one corresponding to the shear fluctuations:
\be \labb{hs} \ddot{H}_{12} +
\dot{H}_{12}\le(3\dot{A} +\frac{\dot{h}}{h}\ri) + H_{12}\frac{\o_h^2}{h^2}=0\,.
\ee

 One crucial difference between (\ref{hb}) and (\ref{hs}) is that, unlike in the case of the shear deformation,
  the bulk deformation  has a mass term even in the hydrodynamic limit $\o_h=0$.  This implies that in general
  there should be a non-trivial flow from the horizon to the boundary
in the sense of the membrane paradigm \cite{LiuIqbal}.  This flow is absent only in the case $X=const$ which
 corresponds to the adiabatic limit \cite{transport}. It is also absent in the Chamblin-Reall solution that
 corresponds to constant $X$, see section \ref{secCR2}.

In the following we apply the formalism developed here to calculate the bulk viscosity in two examples.

\subsection{Numerical results for the holographic-QCD model}

As another non-trivial example, we would like to confirm the agreement between the EO and the GPR
 formula in the improved holographic QCD model of \cite{GK}. The model is based on a single scalar in the bulk theory corresponding to the operator
 $\tr F^2$ in the $SU(N)$ gauge theory. Therefore the deformation in the UV is marginally relevant, hence
 the UV asymptotics is not of the standard asymptotically AdS type, but involve logarithmic
 corrections. In the following we present the results in the variable
\be\labb{defl}
\l = e^{\sqrt{\frac38}\f}.
\ee
The scalar potential is given by,
\be\label{potential}
V(\l)  = -\frac{12}{\ell^2} \left\{ 1 + V_0 \l + V_1 \l^{4/3} \left[\log \left(1 + V_2 \l^{4/3} + V_3 \l^2\right) \right]^{1/2}
\right\} \,,
\ee
The various parameters in (\ref{potential})
\be\labb{pars}
\{V_0,V_1,V_2,V_3\} = \{0.0413\,,\ 14\,,\ 5.3 10^{-9}\,,\ 170\}\,,
\ee
are fixed by in order to fit the UV asymptotics of $SU(N)$ beta-function, the observed latent heat of the
 confinement-deconfinement transition on the lattice and the agreement with the glueball spectrum in the vacuum theory
\cite{GKMN3}.

\begin{figure}
 \begin{center}
\psfrag{elh}{$\lambda_h$}
\psfrag{zs}{$\frac{\zeta}{s}$}
\includegraphics[height=6cm,width=9cm]{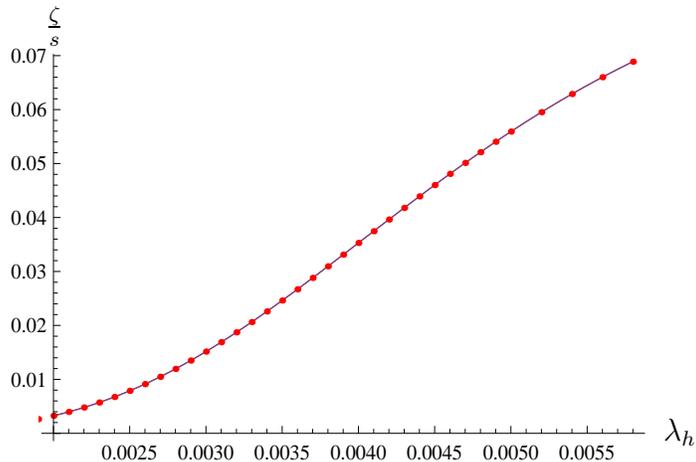}
 \end{center}
 \caption[]{Comparison of the bulk viscosity in the phenomenological QCD model. The solid (blue) curve represents the outcome of the Eling-Oz formula. The (red) dots  are the outcome of Gubser et al's formula.} \label{figpheno1}
 \end{figure}

A straightforward application of the method explained in section \ref{xxx} yields the
 bulk viscosity from the GPR formula \cite{transport}. In the figure \ref{figpheno1} we compare the outcome
  of the two formulae for a range of $\l_h$ that corresponds to the entire range of temperatures above the deconfinement
  transition, $T>T_c$ on the thermodynamically favored and stable big black-hole branch. As seen from this figure the two formulae match to great accuracy.

\addcontentsline{toc}{section}{Acknowledgments}
\section*{Acknowledgments}
We would like to thank Yaron Oz for valuable correspondence and discussions.
Research at Perimeter Institute is
supported by the Government of Canada through Industry Canada and by
the Province of Ontario through the Ministry of Research \&
Innovation. AB gratefully acknowledges further support by an NSERC
Discovery grant and support through the Early Researcher Award
program by the Province of Ontario.
This work was  partially supported by European Union grants FP7-REGPOT-2008-1-CreteHEPCosmo-228644, and PERG07-GA-2010-268246 .

\appendix
\addcontentsline{toc}{section}{Appendices}
\section{$\frac{dA}{d\phi}$ in $\caln=2^*$ plasma\label{A}}

From \cite{ns6}, to leading order in $m_b^2/T^2$ (notice the $\sqrt{24}$ renormalization of the $\a$ to insure
the canonical kinetic term as in \eqref{5dact}),
\begin{equation}
\begin{split}
A(x)=&\ln\dd_3-\frac 14\ \ln \left(2x-x^2\right)+\dd_1^2\ A_1(x)\,,\\
\phi(x)=&\sqrt{24}\ \dd_1\ \a_1(x)\,,
\end{split}
\eqlabel{hightsol}
\end{equation}
where
\begin{equation}
\a_1=\left(2x -x^2\right)^{1/2}\ _2 F_1\left(\ft 12,\ft 12; 1; (1-x)^2\right)\,,
\eqlabel{orko1}
\end{equation}
\begin{equation}
\begin{split}
A_1=&4\int_x^1\ \frac{(z-1)dz}{(2z-z^2)^2}
\left(\gamma_1-\int_z^1dy\left(\frac{\del\a_1}{\del y}
\right)^2\frac{(2y-y^2)^2}{y-1}\right)\,,
\end{split}
\eqlabel{expsol}
\end{equation}
\begin{equation}
\gamma_1=\frac{8-\pi^2}{2\pi^2}\,,\qquad 2\pi T=\dd_3\left(1+\frac{16}{\pi^2}\ \dd_1^2\right)\,,\qquad
\dd_1=-\frac{1}{24\pi}\ \left(\frac{m_b}{T}\right)^2\,.
\eqlabel{gai}
\end{equation}

From \eqref{hightsol}-\eqref{gai} it is easy to deduce that near the horizon, \ie $x\to 1_-$,
\begin{equation}
\begin{split}
\phi=&\phi_h \left(1-\frac 14 (1-x)^2+\calo((1-x)^4)\right)\,,\qquad \phi_h=-\frac{m_b^2}{2\sqrt{6}\pi T^2}\,,\\
A=&\ln(2\pi T)-\frac{m_b^4}{36\pi^4 T^4}+\left(\frac 14-\frac{\gamma_1 m_b^4}{288 \pi^2 T^4}\right) (1-x)^2+\calo((1-x)^4)\,.
\end{split}
\eqlabel{hora1}
\end{equation}
We can now compute $\frac{dA}{d\phi}$ while keeping $m_b$ fixed, see \eqref{apeo} and \eqref{apgu}.

\section{Coefficients $\calh_i$\label{B}}

In the appendix we explicitly show the coefficients of the fluctuation equation (\ref{hx}):

\begin{equation}
\begin{split}
&\calh_1=\frac{(\phi'^2 (x-1)-12 A' (2 (x-1) A'+1)) V_{,\phi}}{(x-1) \phi' V}+\frac{\phi'^2 (x-1)-3 A' (3+8 (x-1) A')}{3(x-1) A'}\,,
\end{split}
\eqlabel{calh1}
\end{equation}
\begin{equation}
\begin{split}
\calh_2=&\frac{ \phi'^2 (x-1)-12 A' (2 (x-1) A'+1)}{6(x-1)^3 \phi' V A'}\ \times
\biggl(2 \phi' V (1-x)+6 V_{,\phi} (1-x) A'\\
&+3 e^{-2 A} \w^2 \phi' A'\biggr)\,.
\end{split}
\eqlabel{calh2}
\end{equation}


\section{\label{cr} Chamblin-Reall backgrounds}
\label{secCR1}

We choose an exponential potential (known also as the Chamblin-Reall geometry), \cite{chr}. In this case
\begin{equation}
V=\calv e^{g \phi}\,,
\eqlabel{potchr}
\end{equation}
with constant $\calv$.

Solving the background equations we find,
\begin{equation}
\begin{split}
A(x)=&\cala-\frac{1}{3g}\ \phi(x)\,,\\
\exp\left(\frac{3 g^2-8}{6 g}\ \phi(x)\right)=&\frac{1}
{\calp(x-1)^2+1-\calp}\,,
\end{split}
\end{equation}
where $\cala, \calp$ are the integration constants, and without the loss of
generality we assumed $\phi(0)=0$.

To leading order in the hydrodynamic limit \eqref{hx} simplifies
dramatically
\begin{equation}
0=H_{11}''+\frac{1}{x-1}\ H_{11}'\,.
\eqlabel{hxcr}
\end{equation}
We outline now the solution of the boundary value problem that we will
use in the more complicated example of the $\caln=2^*$ gauge theory
below.

First, the general solution with the UV boundary condition \eqref{uv1}
is given by
\begin{equation}
H_{11}^b=1+h_{uv}\ \ln(1-x)\,,
\eqlabel{h11uv1}
\end{equation}
with an arbitrary constant $h_{uv}$.
It is straightforward to rewrite \eqref{hxcr} in terms of $y=1-x$,
and find the most general solution satisfying the IR boundary condition
\eqref{ir1}
\begin{equation}
H_{11}^h=h_{ir}\,,
\eqlabel{h11uv}
\end{equation}
with an arbitrary constant $h_{ir}$.
Matching $H_{11}^b$  and $H_{11}^h$ (the value of the function
and the first derivatives) uniquely determines
\begin{equation}
h_{uv}=0\,,\qquad h_{ir}=1\,.
\eqlabel{hs1}
\end{equation}
Thus, much like in \cite{f1}, we conclude that for the Chamblin-Reall model
\begin{equation}
c_{11}^-=1\,.
\eqlabel{c11cr}
\end{equation}

We may also present the results above in the language of phase variables,
\label{secCR2} (\ref{XY}).
The Chamblin-Reall solution is given by (see Appendix J of \cite{GKMN2}),
\begin{equation}\label{CRXY}
    X = -\sqrt{\frac38} g\,, \qquad Y = \frac{1-X^2}{e^{\a(\f_h-\f)}-1}\,,
\end{equation}
where we defined the constant,
\begin{equation}\label{adef}
    \a = -\sqrt{\frac23}\frac{(1-X^2)}{X}\,.
\end{equation}
We note that for consistency of thermodynamics $X^2<1$, otherwise the black-hole solution has
 negative specific heat, hence corresponds to a small black-hole \cite{GKMN2}. One finds the following metric functions
in the variable $\f$:
\bea
A(\f) &=& A_0 + \frac{1}{\sqrt{6}X}\f\,, \label{ACR}\\
h(\f) &=& 1- e^{\a(\f-\f_h)}\,.\label{hCR}
\eea
One distinguishing fact about the above solution is that the scale function $A(\f)$ is independent of the temperature $\f_h$.

 Before making this connection however, let us provide a simple proof---closely related to the one given in  section \ref{secCR1}. The fluctuation equation
 (\ref{hwXY}) simplifies drastically as the coefficient $d(\f)$ in (\ref{df}) vanishes for $\w =  0$. This means in particular that there is no flow from the horizon
 to the boundary in the sense of the membrane paradigm, see e.g. \cite{LiuIqbal} for the bulk-viscosity in the case of Chamblin-Reall backgrounds.

 The proof that $|c_{11}^-|(\f_h)= 1$ in this case is already given in Appendix B of \cite{transport}, that we review here. When, the coefficient $d(\f)$ in (\ref{df}) vanishes in the $\w=0$ equation, the solution to $H_{11}$ is simply  given by,
\be\labb{fe2} H_{11}(\f) = 1
+ C \int_{-\f_0}^{\f} dt\,\, e^{\int_{-\f_0}^{t} c(t)}\,, \ee
where the function $c(\f)$ is given by (\ref{cf}) and we used the boundary condition $H_{11}(\f_0)=1$.
The second integration constant $C$ is determined by the second boundary
condition that $H_{11}(\f)$ is regular at the horizon\cite{f1}. On the other hand, the function
$c(\f)$ in (\ref{cf}) is positive definite because $X<0$, $Y>0$, $X^2<1$, and the term inside the brackets is given by $-4/3g$, hence negative.  Therefore the only way to guarantee regularity at the horizon is to set $C=0$,
hence $H_{11} = 1$ for all values of $\l$ in the limit, in particular $|c_{11}^-|$ is 1.

\end{document}